\documentclass[aps,superscriptaddress,amsmath,preprint]{revtex4}
\usepackage{verbatim}
\usepackage{amsmath,amssymb,amsthm}
\usepackage{dcolumn}
\usepackage{bm}
\usepackage{graphicx}
\DeclareMathOperator{\sign}{sign}

\begin{document}

%\title{Semi-metallic graphene based diode}
\title{Klein collimation by rippled graphene superlattice}
\author{M. Pudlak}
 \affiliation{Institute of Experimental
Physics,  04001 Kosice, Slovakia}
\author{R.G. Nazmitdinov}
\email{rashid@theor.jinr.ru}
\affiliation{Bogoliubov Laboratory of Theoretical Physics,
Joint Institute for Nuclear Research, 141980 Dubna,
Moscow region, Russia}
\affiliation{Dubna State University, 141982 Dubna, Moscow region, Russia}

\begin{abstract}
The hybridization of $\sigma$ and $\pi$ orbitals of carbon atoms
in graphene depends on the surface curvature.
Considering a single junction between flat and rippled
graphene  subsystems, it is found an accumulation of charge
 in the rippled subsystem due to Klein penetration phenomenon
 that gives rise to n-p junction.
Using this fact, we show that the momentum
distribution of  electrons in ballisitically propagating beam can be selective without a waveguide, or
external electric, and/or  magnetic fields  in graphene strip under experimentally feasible
one-dimensional periodic potential.
Such a potential is created with the aid of superlattice
that consists of periodically repeated graphene pieces with
different hybridizations of carbon orbits, produced by variation of the graphene surface curvature.
The charge redistribution and selected transmission of electrons, caused by the superlattice,
allows to control the electron focusing
in the considered system by simply changing the  element properties in the superlattice.
\end{abstract}

\date{\today}

\maketitle

\section{Introduction}
There is an enormous  experimental and theoretical activity devoted to
graphene and graphene based devices.
Indeed, a graphene being a zero-gap semiconductor yields
exceptionally high mobility of charged carriers.
However, the inability to control this mobility
in a graphene is a supreme concern of nanoelectronics.
Nevertheless, unique properties of graphene nanostuctures, discussed below, offer a promicing
approach in this field.

The low-energy spectrum of graphene  is quite well described  theoretically in the effective mass
approximation by the linear energy dispersion, which is the same as Weyl's equation for massless neutrino \cite{Ando05}. This description has been proved experimentally, for example,
by the observation of a relativistic analogue of the integer Hall effect (e.g., \cite{13,14}).
The linear dispersion  is explained as a consequence of graphene crystal structure that consists
of two equivalent carbon sublattices. This fact allows to introduce graphene quasiparticles with
different pseudospin quantum numbers associated with  corresponding sublattices. As a result,
such quasiparticles  are expected to behave differently from those in conventional
metals and semiconductors \cite{kat1}. It was shown in Ref.\cite{Falko} that the conservation
of the pseudospin forbids strictly charged carrier backscattering in a graphene
monolayer with  electrostatic potential scattering that mimics the n-p junction.
The barrier always remains perfectly transparent for the normal incidence of electrons,
while the transmission decreases for other angles.  By virtue of this fact, electron focusing
analogous to optical effects that occur in negative refractive index material is predicted \cite{ch1}.
In fact, it was demonstrated experimentally that: i) turning  carrier density
in graphene sheet by means of electrical gates \cite{huard}; ii) using electrostatic dopping
from buried gates \cite{sutar}, or iii) transverse magnetic focusing \cite{mag}, - it is possible to obtain angle-dependent carrier transmission in graphene n-p junction. These results confirm evidently  that, indeed,
electron transport through graphene n-p junction has much  resemblance to light rays crossing a
boundary between materials with different optical index.  All these phenomena are founded on unimpeded
Klein tunneling penetration \cite{klein} through gate potential barriers, that is used recently
to create a graphene transistor on tunable fermion optics \cite{switch}.

It is noteworthy to mention that above discussed results are based on assumption of use
\textit{external electrical or magnetic}  accessories to control the focusing of electron flow.
We recall, however, that graphene sheets are not perfectly flat, and
ripples are considered as most natural sources that might be used
to control the electron mobility as well.
Moreover, by means of the DFT and molecular dynamics simulations it is shown that graphene
demonstrates extraordinary stretchability, up to about 20--30\%, without being
damaged \cite{kumar}. The amplitude and the orientation of the unidirectional ripples
can be controlled with the aid of the applied strain \cite{jul1}.
And further, it is shown that using the hydrogenation it is possible
to induce periodic ripples with various thermal conductivity \cite{jul2}.

The effect of the
corrugations in graphene on the electronic structure and density of
states was evidently demonstrated  in Ref.\cite{Voz}.
It is predicted that ripples  could create in graphene:
i) electron scattering, caused by
the change in nearest-neighbor hopping parameters by the curvature
\cite{Kat,Guinea}:
ii) an electrostatic potential \cite{PPCF,Allain};
or iii) a chiral transport \cite{PPN1} due to a spin-orbit interaction induced by
the surface curvature \cite{Ando,PPN}.
Furthemore, one-dimensional (1D) nanoscale periodic ripples  could
generate a periodic electronic graphene superlattice \cite{Wei,Bai}.

Note, that typical transition lengths for n-p junction are less 100 nm (e.g., \cite{huard}).
A ballistic transport model is
sufficient for the study of physics n-p junction devices \cite{low}.
It is appropriate at this point to mention a natural way to control
the dispersion of the ballistic electron beam in graphene based
systems. It was shown in Refs.\cite{Pudlak,Pudlak1} that the
hybridization of $\pi$ electron orbital of carbon atom depends on the
hybridization of $\sigma$ orbitals. As a matter of fact, the hybridization is different in
a flat and a corrugated graphene.
The purpose of the present paper is to exploit this fact
and suggest the novel n-p junction based on
different hybridizations of carbon orbits, produced by variation of the graphene surface curvature.
Considering the superlattice
that consists of periodically repeated graphene pieces with
different hybridizations of carbon orbits,
we will demonstrate its high angle-dependent selectivity
of the  transmitted ballistic electrons. This selectivity  allows
the electron  focusing at low-energy physics of graphene
without any additional electrical or magnetic sources, simply
by element settings  in the superlattice.

\section{The mechanism of hybridization in a curved graphene}

Let us specify the mechanism of
hybridization of $\pi$ and $\sigma$ orbitals in the flat and the curved
graphene systems.
For the sake of discussion, we recapitulate the basic results
for the flat graphene in the effective mass approximation (e.g., \cite{SDD}).

We consider the Hamiltonian
for the $K$ point  (similar approach can be applied for
$K^{'}$ point). It depends on two operators
$\hat{k}_{x}=-i\frac{\partial}{\partial x}$,
$\hat{k}_{y}=-i\frac{\partial}{\partial y}$, and
yields the equation for the envelope function of the flat graphene \cite{SDD}
\begin{equation}
\label{1} \left(\begin{array}{cc}\varepsilon_{2p}&\gamma
(\hat{k}_{x}-i\hat{k}_{y})\\\gamma
(\hat{k}_{x}+i\hat{k}_{y})&\varepsilon_{2p}
\end{array}\right)\left(\begin{array}{c}
F^{K}_{A}\\F^{K}_{B}\end{array}\right)=E
\left(\begin{array}{c}F^{K}_{A}\\F^{K}_{B}\end{array}\right)\,.
\end{equation}
Here, the parameter $\gamma =\sqrt{3}\gamma_{0}a/2$ depends on  the length of the
primitive translation vector $a=\sqrt{3}d\simeq 2.46 A^{\circ}$, with
$d$ being the distance between atoms in the unit cell, and it is
assumed that $\gamma_{0}\approx 3$ eV. The energy
$\varepsilon_{2p}=\langle 2p_{z}|\verb"H"|2p_{z} \rangle$
is the energy of $2p_{z}$-orbitals of carbon atoms in the flat graphene,
directed perpendicular to the graphene surface;
$\verb"H"$ is the tight-binding Hamiltonian of the graphene.
The solution of Eq. (\ref{1}) determines the wave function
\begin{equation}
\label{wff}
F(x,y)=e^{ik_{x}x}e^{ik_{y}y}\frac{1}{\sqrt{2}}\left(\begin{array}{c}s
e^{-i\varphi}\\1\end{array}\right),\quad
e^{-i\varphi}=(k_{x}-ik_{y})/\sqrt{k_{x}^{2}+k_{y}^{2}}\,,
\end{equation}
and the energy
\begin{equation}
E=\varepsilon_{2p}+s \gamma \sqrt{k_{x}^{2}+k_{y}^{2}}\,.
\end{equation}
Here, the sign $s=-1(+1)$ is associated with the valence (conductance) band.
In the flat graphene we have the following hybridization of $\pi$ and
$\sigma$ orbitals:
\begin{eqnarray}
&|\pi\rangle = |2p_{z}\rangle\,,\\
&|\sigma_{1}\rangle = \frac{1}{\sqrt{3}}|2s\rangle
+\sqrt{\frac{2}{3}}|2p_{y}\rangle\,,\\
&|\sigma_{2}\rangle = \frac{1}{\sqrt{3}}|2s\rangle +
\sqrt{\frac{2}{3}}\left(\frac{\sqrt{3}}{2}|2p_{x}\rangle
-\frac{1}{2}|2p_{y}\rangle \right)\,,\\
&|\sigma_{3}\rangle = \frac{1}{\sqrt{3}}|2s\rangle -
\sqrt{\frac{2}{3}}\left(\frac{\sqrt{3}}{2}|2p_{x}\rangle
+\frac{1}{2}|2p_{y}\rangle \right)\,.
\end{eqnarray}

Let us discuss the hybridization of
$\sigma$ and $\pi$ orbitals in the graphene with nonzero curvature.
The $\sigma$ orbitals create the bonds between carbon atoms, while the
$\pi$ orbitals determine the electronic properties of the graphene.
\begin{figure}[hbt]
\centerline{\includegraphics{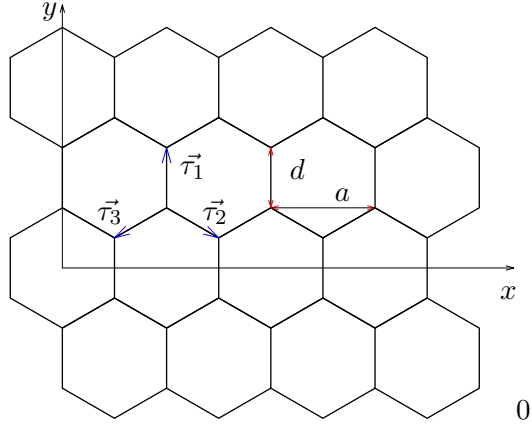}\unitlength=1mm\begin{picture}(0,0)(0,0)
\put(-62,53){\makebox(0,0)[b]{$y$}}
\put(-1,16){\makebox(0,0)[b]{$x$}}
\put(-43,32){\makebox(0,0)[b]{$\vec{\tau_{1}}$}}
\put(-40,26){\makebox(0,0)[b]{$\vec{\tau_{2}}$}}
\put(-54,26){\makebox(0,0)[b]{$\vec{\tau_{3}}$}}0
\put(-31,32){\makebox(0,0)[b]{$d$}}
\put(-25,29){\makebox(0,0)[b]{$a$}}
\end{picture}}
\caption{Graphene lattice. It is assumed that the graphene sheet is wrapped
into the tube form with the symmetry axis in the $y$ direction.}
\label{grafdraw}
\end{figure}

For the sake of illustration we  consider a zig-zag nanotube
(see Fig.\ref{grafdraw}). For the curved graphene (the arc, characterised by the
radius $R$) we obtain the space coordinates of the three
nearest-neighbor vectors $\vec{\tau}_{i}$ in the following form:
\begin{eqnarray}
&\vec{\tau}_{1}=d(0,1,0) \,,\\
&\vec{\tau}_{2}=d(\frac{\sqrt{3}}{2}\cos\alpha
,-\frac{1}{2},-\frac{\sqrt{3}}{2}\sin\alpha )\,,\\
&\vec{\tau}_{3}=d(-\frac{\sqrt{3}}{2}\cos\alpha
,-\frac{1}{2},-\frac{\sqrt{3}}{2}\sin\alpha )\,,
\end{eqnarray}
where $\sin\alpha =a/4R$. At the limit $R\rightarrow \infty$,  the vectors
$\vec{\tau}_{i}$ transform to those of the flat graphene. Evidently, the
$\sigma_{i}$ -orbitals are determined by the vectors $\vec{\tau}_{i}$.
As a result, the $\sigma_{i}$ and $\pi$ orbitals can be expressed
as follows
\begin{eqnarray}
&|\sigma_{1}\rangle = c_{1}|2s\rangle
+\sqrt{1-c_{1}^{2}}|2p_{y}\rangle\,,\\
&|\sigma_{2}\rangle = c_{2}|2s\rangle +
\sqrt{1-c_{2}^{2}}\left(\frac{\sqrt{3}}{2}\cos\alpha |2p_{x}\rangle
-\frac{1}{2}|2p_{y}\rangle-\frac{\sqrt{3}}{2}\sin\alpha
|2p_{z}\rangle \right)\,,\\
&|\sigma_{3}\rangle = c_{3}|2s\rangle +
\sqrt{1-c_{3}^{2}}\left(-\frac{\sqrt{3}}{2}\cos\alpha |2p_{x}\rangle
-\frac{1}{2}|2p_{y}\rangle
-\frac{\sqrt{3}}{2}\sin\alpha|2p_{z}\rangle \right)\,,\\
&|\pi\rangle = d_{1}|2s\rangle +d_{2}|2p_{x}\rangle
+d_{3}|2p_{y}\rangle +d_{4}|2p_{z}\rangle\,.
\end{eqnarray}

With the aid of the orthonormality
conditions $\langle\sigma_{i}|\sigma_{j}\rangle =\delta_{ij}$,
$\langle \pi|\sigma_{j}\rangle =0$, and $\langle \pi|\pi\rangle =1$,
we determine the parameters $\{c_{k},d_{l}\}$ and obtain
the following expressions for the $\pi$ and $\sigma$ orbitals in
the lowest order of the ratio $a/R$:
\begin{eqnarray}
\label{1f}
&|\pi\rangle \approx |2p_{z}\rangle +\frac{a}{2\sqrt{6}R}|2s\rangle
+\frac{a}{4\sqrt{3}R}|2p_{y}\rangle\,,\\
\label{2f}
&|\sigma_{1}\rangle = \frac{1}{\sqrt{3}}|2s\rangle
+\sqrt{\frac{2}{3}}|2p_{y}\rangle\,,\\
\label{3f}
&|\sigma_{2}\rangle = \frac{1}{\sqrt{3}}|2s\rangle +
\sqrt{\frac{2}{3}}\left(\frac{\sqrt{3}}{2}|2p_{x}\rangle
-\frac{1}{2}|2p_{y}\rangle-\frac{\sqrt{3}a}{8R}|2p_{z}\rangle
\right)\,,\\
\label{4f}
&|\sigma_{3}\rangle = \frac{1}{\sqrt{3}}|2s\rangle -
\sqrt{\frac{2}{3}}\left(\frac{\sqrt{3}}{2}|2p_{x}\rangle
+\frac{1}{2}|2p_{y}\rangle +\frac{\sqrt{3}a}{8R}|2p_{z}\rangle
\right)\,.
\end{eqnarray}
The $\pi$ orbitals are the same for the zig-zag and armchair nanotubes in
the lowest order of $a/R$. They are used to create the Bloch function
in the tight-binding approximation. As a result, we obtain  the following
$\pi$ orbital energy of the curved  graphene surface of radius $R$
\begin{eqnarray}
\varepsilon_{\pi} &=&\langle \pi|\verb"H"|\pi \rangle =\langle
2p_{z}|\verb"H"|2p_{z} \rangle +\frac{1}{24}\left(
\frac{a}{R}\right)^{2}\langle 2s|\verb"H"|2s \rangle
+\frac{1}{48}\left( \frac{a}{R}\right)^{2}\langle
2p_{y}|\verb"H"|2p_{y} \rangle\,=\nonumber\\
  &=&  \varepsilon_{2p}+ \alpha
\left(\frac{a}{R}\right)^{2}\,,
\quad \alpha = \frac{1}{24}\langle s|\verb"H"|s \rangle +
\frac{1}{48}\langle p_{y}|\verb"H"|p_{y} \rangle\,.
\label{2}
\end{eqnarray}
Note, that the orbitals $2p_{y,z}, 2s$ are localized on the
same carbon atom and contribute to the $\pi$ orbital energy \cite{SDD},
while there is no such a contribution from the nondiagonal matrix
elements.
As a result, we obtain that the energy of the curved graphene consists of
the energy of the flat graphene $\varepsilon_{2p}$, and  the energy of the
$2s$, $2p_y$ orbitals brought about by the curvature (see also \cite{Pudlak}).

Using the numerical values for the energies of the $|s\rangle$ and $|p_{y}\rangle$
orbitals  of the carbon atom
$\langle s|\verb"H"|s \rangle =-12$eV, $\langle p_{y}|\verb"H"|p_{y}
\rangle =-4$eV (e.g., \cite{Lomer}),
we obtain for the parameter $\alpha \simeq -0.58$eV. Thus, the energy difference
between the $\pi$ orbitals of the curved and flat graphene is
\begin{equation}
\varepsilon_{2p}-\varepsilon_{\pi}=\Delta \varepsilon =|\alpha |
\left(\frac{a}{R}\right)^{2}\approx
0.58\left(\frac{a}{R}\right)^{2}eV\,.
\label{endif}
\end{equation}
In  the curved graphene
the effective mass Hamiltonian, Eq.(\ref{1}), transforms to the form
\begin{equation}
\hat{H}=\varepsilon_{\pi}\sigma_{0}+\gamma
(\hat{k}_{x}\sigma_{x}+\hat{k}_{y}\sigma_{y})\,. \label{hr}
\end{equation}
Here, $\sigma_{x},\sigma_{y}$ are the Pauli matrices and $\sigma_{0}$ is
the unity matrix. Solving  the Schr\"{o}dinger equation with the Hamiltonian (\ref{hr}),
we obtain the wave function for the curved region
\begin{equation}
\label{wfr}
F(x,y)=e^{i\kappa_{x}x}e^{ik_{y}y}\frac{1}{\sqrt{2}}\left(\begin{array}{c}s
e^{-i\chi}\\1\end{array}\right)\,,\quad
e^{-i\chi}=(\kappa_{x}-ik_{y})/\sqrt{\kappa_{x}^{2}+k_{y}^{2}}\,,
\end{equation}
with eigenvalue $E=\varepsilon_{\pi} +s
\gamma\sqrt{\kappa_{x}^{2}+k_{y}^{2}}$.

The difference between $\varepsilon_{\pi}$ (curved region) and
$\varepsilon_{2p}$ (flat region) is important when the systems with
different surface curvature are coupled. Hereafter, for the sake of
simplicity we assume that $\varepsilon_{\pi}=0$.

\section{The hybrid graphene system}
\subsection{Simple junction}
\begin{figure}{}
\centerline{\includegraphics{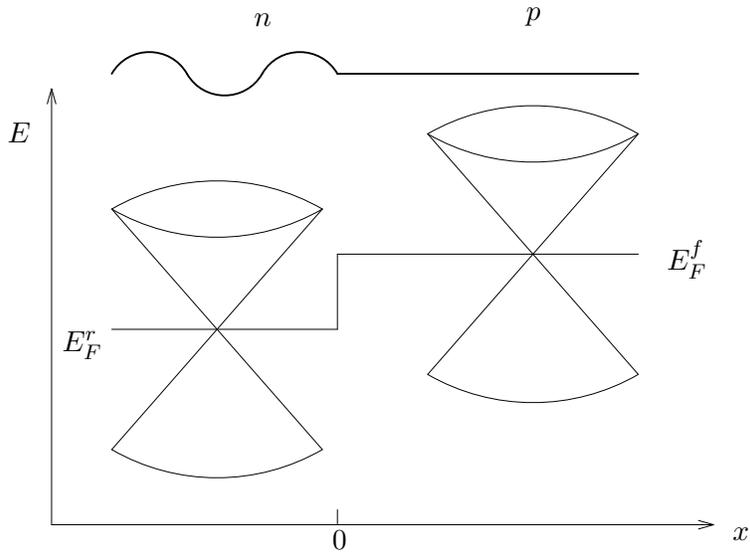}\unitlength=1.2mm\begin{picture}(0,0)(0,0)
\put(-77,43){\makebox(0,0)[b]{$E$}}
\put(-50,56){\makebox(0,0)[b]{$n$}}
\put(-20,56){\makebox(0,0)[b]{$p$}}
\put(-70,19){\makebox(0,0)[b]{$E_{F}^{r}$}}
\put(-3,28){\makebox(0,0)[b]{$E_{F}^{f}$}}
\put(-41.5,-2){\makebox(0,0)[b]{$0$}}
\put(3,-1){\makebox(0,0)[b]{$x$}}
\end{picture}}
\caption{The sketch of the flat-rippled graphene junction.
Here, $E_{F}^{f}$ is the Fermi energy of the flat graphene
and $E_{F}^{r}$ is the Fermi energy of the rippled graphene.}
\label{rippled}
\end{figure}

Hereafter, we consider a wide enough graphene sheet $W\gg M$,
where $W$ and $M$ being, respectively, as the width along the y
axis and the length along x axis of the graphene sheet.
It means that we keep the translational invariance along the y axis
and neglect the edge effects.
Due to different hybridization mechanisms, the Fermi energy of the flat
graphene is higher than the Fermi energy of the corrugated graphene
(see Sec.II). Let us consider the system (see Fig.\ref{rippled})
that consists of the flat graphene piece ($x>0$) connected to the
corrugated graphene piece ($x \leq 0$). In this case the corrugated
graphene is modelled by consistently connected arc (with the radius
R) and inverted arc (with the same radius) pieces etc. For the sake
of convenience,  we introduce the notation ${\cal R} $(${\cal F})$
for a rippled (flat) graphene system.

In the combined system (${\cal R+F}$)
electrons flow from the ${\cal F}$ subsystem
to the ${\cal R}$ subsystem. The flow stops
once the potential energy difference between two sides of the
junction is equal in magnitude and opposite in sign to the
difference between two local Fermi levels, similarly to the bimetal
interface \cite{Yan}.
The common Fermi level of two subsystems  is determined
as
\begin{equation}
E_{F}= \frac{1}{2}(E_{F}^{f}+E_{F}^{r})=E_{F}^{r}+
\frac{|\alpha|}{2} \left(\frac{a}{R}\right)^{2}. \label{ef}
\end{equation}
Here,  the local Fermi energy of the flat region is $E_{F}^{f}\equiv\varepsilon_{2p}$, while
the local Fermi energy of rippled region is $E_{F}^{r}\equiv \varepsilon_{\pi}\equiv0$ (see Sec.II).
With the aid of the definition of  density of states in the graphene (see Ref.\cite{Wallace})
\begin{equation}
N(E)=g_{s}\frac{2|E-E_{F}|}{3\pi \gamma_{0}^{2}a^{2}}
\end{equation}
and Eq.(\ref{ef}), we can define the number of electrons moving to the
${\cal R}$ region
\begin{equation}
n=g_{s}\int_{E_{F}^{r}}^{E_{F}}\frac{2|E-E_{F}^{r}|}{3\pi
\gamma_{0}^{2}a^{2}}dE =g_{s}\int_{0}^{\frac{|\alpha |}{2}
\left(\frac{a}{R}\right)^{2}}\frac{2EdE}{3\pi \gamma_{0}^{2}a^{2}}
=\frac{1}{3\pi
a^{2}}\left(\frac{\alpha}{\gamma_{0}}\right)^{2}\left(\frac{a}{R}\right)^{4}\,.
\end{equation}
As a result, taking into account  the degeneracy value $g_{s}=4$ (spin and valley),
$\gamma_{0}=3$eV and $a=2.46{\AA}$,
the electron density profile in the ${\cal R}$ region, determined by the expression
\begin{equation}
n(R)=6.6\times 10^{12}\left(\frac{2.46}{R}\right)^{4}cm^{-2}\,,
\end{equation}
yields the density
$n\sim 1.6\times 10^{12}cm^{-2}$ at the $R=3.5{\AA}$.
In other words, there is the extra charge $\Delta Q=e n$ per area in the ${\cal R}$ region
and the lack of this charge in the ${\cal F}$ region ($e$ is
the charge of the electron). This situation implies the creation of the n-p junction
due to the different hybridization mechanisms. Could we use this fact ?
The answer on this question is addressed below.

\subsection{The superlattice effect}
Thus, combining two subsystems, we have  created a square (sharp) potential step of the height
$V_0=|\alpha|\left(\frac{a}{R}\right)^{2}$ on which an electron of
energy $E=E_F>0$ (Eq.\ref{ef}) is incident.
As it was stressed in Ref.\cite{Allain}, in this situation there is
an evident analogy with the optical system, when a light beam going through
a discontinuity between two transparent media.
Evidently, however, that it is quite difficult to arrange experimentally
a sharp potential step in graphene based systems. Most likely n-p junction
is expected to be atomically  smooth  (e.g., Ref. \cite{mag}).

\begin{figure}[b]
\centerline{\includegraphics{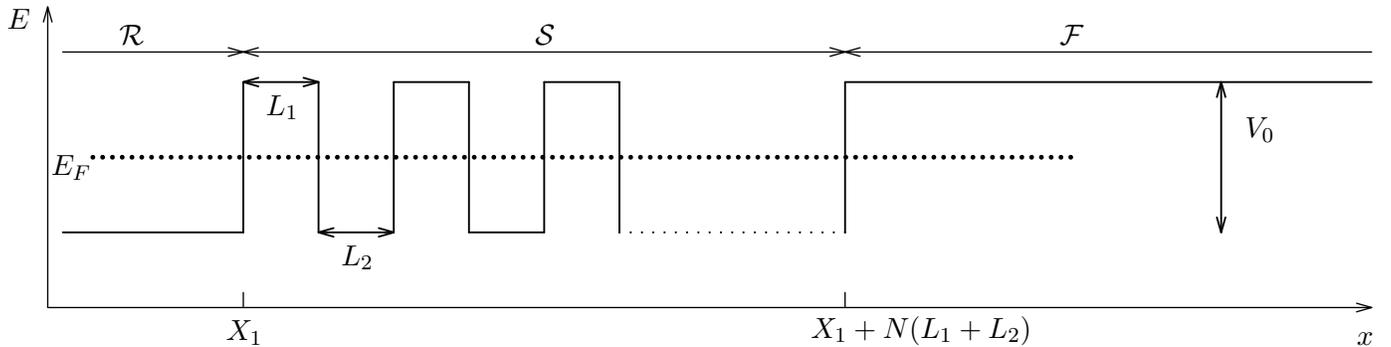}\unitlength=1mm\begin{picture}(0,0)(0,0)
\put(-15,23){\makebox(0,0)[b]{$V_0$}}
\put(-173,18){\makebox(0,0)[b]{$E_F$}}
\put(-145,26){\makebox(0,0)[b]{$L_1$}}
\put(-135,6){\makebox(0,0)[b]{$L_2$}}
\put(-150,-4){\makebox(0,0)[b]{$X_1$}}
\put(-60,-4){\makebox(0,0)[b]{$X_1+N(L_1+L_2)$}}
\put(-1,-4){\makebox(0,0)[b]{$x$}}
\put(-165,36){\makebox(0,0)[b]{${\cal R}$}}
\put(-110,36){\makebox(0,0)[b]{${\cal S}$}}
\put(-40,36){\makebox(0,0)[b]{${\cal F}$}}
\put(-180,38){\makebox(0,0)[b]{$E$}}
\end{picture}}
\caption{The hybrid graphene system ${\cal R+S+F}$. The interface (${\cal S}$) is created between ${\cal R}$ and
${\cal F}$ subsystems. The  superlattice (${\cal S}$) contains $N$ units. Each unit  consists
of the rippled and flat sections. The flat section has length $L_{1}$
and the rippled section (arc) has length $L_{2}$. }
\label{ripplediod}
\end{figure}
To model such a situation in our case,
we consider the hybrid  graphene system that consists of ${\cal
R}$+${\cal S}$+${\cal F}$ regions. We introduce the notation ${\cal
S}$ for the semi-rippled subsystem (see
Fig.\ref{ripplediod}) that consists of N units (superlattice)
with the folowing structure of one unit. It contains the flat and curved (the arc) regions with  lengths
$L_{1}$ and  $L_{2}$, respectively.
In this case we are faced with the
phenomenon of the Klein tunneling (e.g., \cite{Allain} and references therein)
in this hybrid system. To
calculate the Klein tunneling through the superlattice (${\cal S}$ region),
we consider the wave function in the rippled region
$[-\infty \leq x < X_{1}, |y|<W]$ in the form
\begin{equation}
\Psi(x,y)=\left\{e^{i\kappa_{x}x}\frac{1}{\sqrt{2}}\left(\begin{array}{c}e^{-i\chi}\\1\end{array}\right)
+r
e^{-i\kappa_{x}x}\frac{1}{\sqrt{2}}\left(\begin{array}{c}-e^{i\chi}\\1\end{array}\right)\right\}e^{ik_{y}y}\,.
\end{equation}
For the first flat sector of the ${\cal S}$ region
$[X_{1}\leq x < X_{1}+L_{1},  |y|<W]$ we have
\begin{equation}
\Psi(x,y)=\left\{\alpha_{1}e^{ik_{x}x}\frac{1}{\sqrt{2}}
\left(\begin{array}{c}-e^{-i\varphi}\\1\end{array}\right)+
\beta_{1}e^{-ik_{x}x}\frac{1}{\sqrt{2}}\left(\begin{array}{c}e^{i\varphi}\\1\end{array}
\right)\right\}e^{ik_{y}y}\,,
\end{equation}
and for the first rippled sector of the ${\cal S}$ region
$[X_{1}+L_{1} \leq x < X_{1}+L_{1}+L_{2},  |y|<W]$,
we define the wave function in the form
\begin{equation}
\Psi(x,y)=\left\{\gamma_{1}e^{i\kappa_{x}x}\frac{1}{\sqrt{2}}\left(\begin{array}{c}e^{-i\chi}\\1\end{array}\right)
+\delta_{1}
e^{-i\kappa_{x}x}\frac{1}{\sqrt{2}}\left(\begin{array}{c}-e^{i\chi}\\1\end{array}\right)\right\}e^{ik_{y}y}\,,
\end{equation}
and so on. The unknown coefficients
$\alpha_{i},\beta_{i},\gamma_{i},\delta_{i}$ are obtained from the
continuity conditions on the boundary. For the ${\cal F}$ region
$[X_{1}+N(L_{1}+L_{2}) \leq x < M,  |y|<W]$, we have
\begin{equation}
\Psi (x,y)=t
e^{-ik_{x}x}\frac{1}{\sqrt{2}}\left(\begin{array}{c}e^{i\varphi}\\1
\end{array}
\right)e^{ik_{y}y}\,.
\end{equation}
We assume that electron moves with the kinetic energy
$E=E_F\equiv V_{0}/2$, where the barrier height
$V_{0}=|\alpha| \left(\frac{a}{R}\right)^{2}$. At this special case
$k_{x}=\kappa_{x}=k$, and, therefore, $\varphi=\chi$. Using the continuity
conditions on the boundaries, we obtain the following equations for
the transmission coefficient $t$ and reflection coefficient $r$

\begin{equation}
\label{51}
\left(\begin{array}{c}1\\r\end{array}\right)=\left(\begin{array}{cc}A_{11}&A_{12}\\A_{21}&A_{22}
\end{array}\right)^{N}D\left(\begin{array}{c}
0\\t\end{array}\right)\,,
\end{equation}
where
\begin{equation}
\label{51a}
D=\frac{1}{\cos\varphi}\left(\begin{array}{cc}i\sin\varphi&e^{i\varphi}\\e^{-i\varphi}&-i\sin\varphi
\end{array}\right)\,,
\end{equation}
\begin{equation}
A_{11}=e^{-ikL_{2}}\left(\cos k L_{1}+i\sin k L_{1}\frac{1+\sin^{2}\varphi}{\cos^{2}\varphi}\right)
=A_{22}^{*}\,,
\end{equation}
\begin{equation}
A_{12}=2e^{i\varphi}e^{ik L_{2}}\sin k L_{1}\frac{\sin\varphi}{\cos^{2}\varphi}=A_{21}^{*}\,.
\end{equation}
Taking into account that there is the unitary transformation $U$ which diagonalizes the
matrix A, i.e.,
\begin{equation}
U^{-1}AU=\left(\begin{array}{cc}\lambda_1&0\\0&\lambda_{2}
\end{array}\right)\,,
\end{equation}
we can introduce the following notations
\begin{equation}
\label{52} \left(\begin{array}{cc}A_{11}&A_{12}\\A_{21}&A_{22}
\end{array}\right)^{N}=\left(\begin{array}{cc}N_{11}&N_{12}\\N_{21}&N_{22}
\end{array}\right)
\end{equation}
with the following elements:
\begin{equation}
N_{11}=\frac{A_{11}(\lambda_{1}^{N}-\lambda_{2}^{N})+\lambda_{2}^{N-1}-
\lambda_{1}^{N-1}}{\lambda_{1}-\lambda_{2}}=N_{22}^{*}\,,
\end{equation}
\begin{equation}
N_{12}=A_{12}A_{N}=N_{21}^{*}\,,\quad
A_{N}=\frac{\lambda_{2}^{N}-\lambda_{1}^{N}}{\lambda_{2}-\lambda_{1}}\,.
\end{equation}
This trick determines the eigenvalues
\begin{eqnarray}
\label{l12}
&&\lambda_{1,2}=a\pm\sqrt{a^2-1}\,,\quad a=(A_{11}+A_{22})/2\,,\\
\label{23}
&&a=\cos[k(L_{1}-L_{2})]+2\sin(kL_{1})\sin(kL_{2})\frac{\sin^{2}\varphi}{\cos^{2}\varphi}\,.
\end{eqnarray}

As a result we can calculate analytically the electron transmission probability
across the interfaces as
\begin{equation}
\label{trans}
T_{N}(k_{y})=|t|^{2}=\frac{\cos^{2}\varphi}{1+4\frac{\sin^{2}\varphi}{\cos^{2}\varphi}\sin(kL_{1})A_{N}\Pi}\,,
\end{equation}
where
\begin{equation}
\Pi =
A_{N-1}\sin(kL_{2})+A_{N}\left(\sin(kL_{1})+\sin[k(2L_{2}-L_{1})]+
4\frac{\sin(kL_{1})\sin^{2}(kL_{2})\sin^{2}\varphi}{\cos^{2}\varphi}\right)\,,
\end{equation}
\begin{equation}
\sin^{2}\varphi =\frac{k_{y}^{2}}{k_{F}^{2}}\,,\quad k =\sqrt{k_{F}^{2}-k_{y}^{2}}\,.
\end{equation}
\begin{figure}[b]
\includegraphics[height=6cm,clip=]{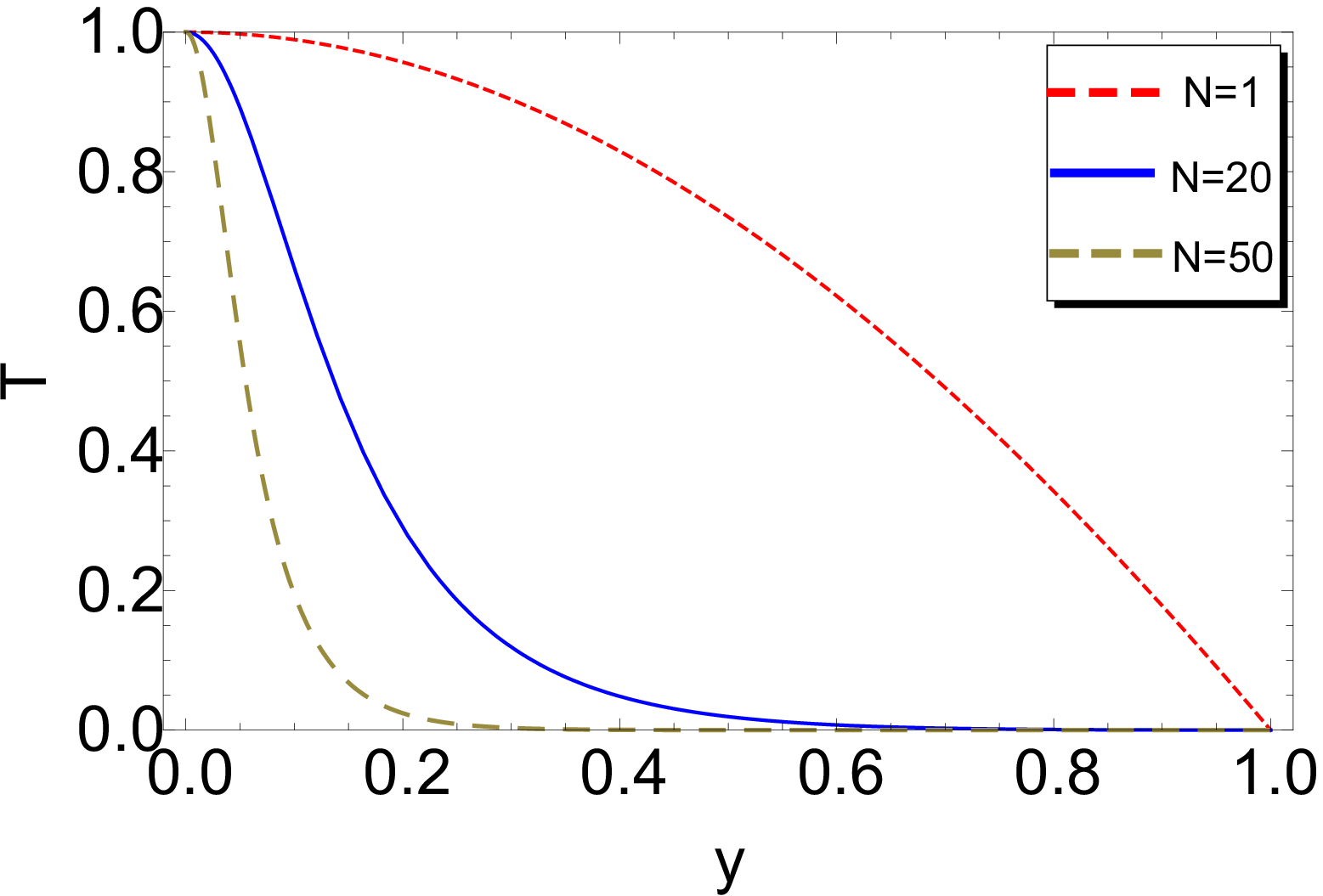}\begin{picture}(0,0)(0,0)
\put(-65,130){\makebox(0,0)[b]{$(a)$}}
\end{picture}\
\includegraphics[height=6cm,clip=]{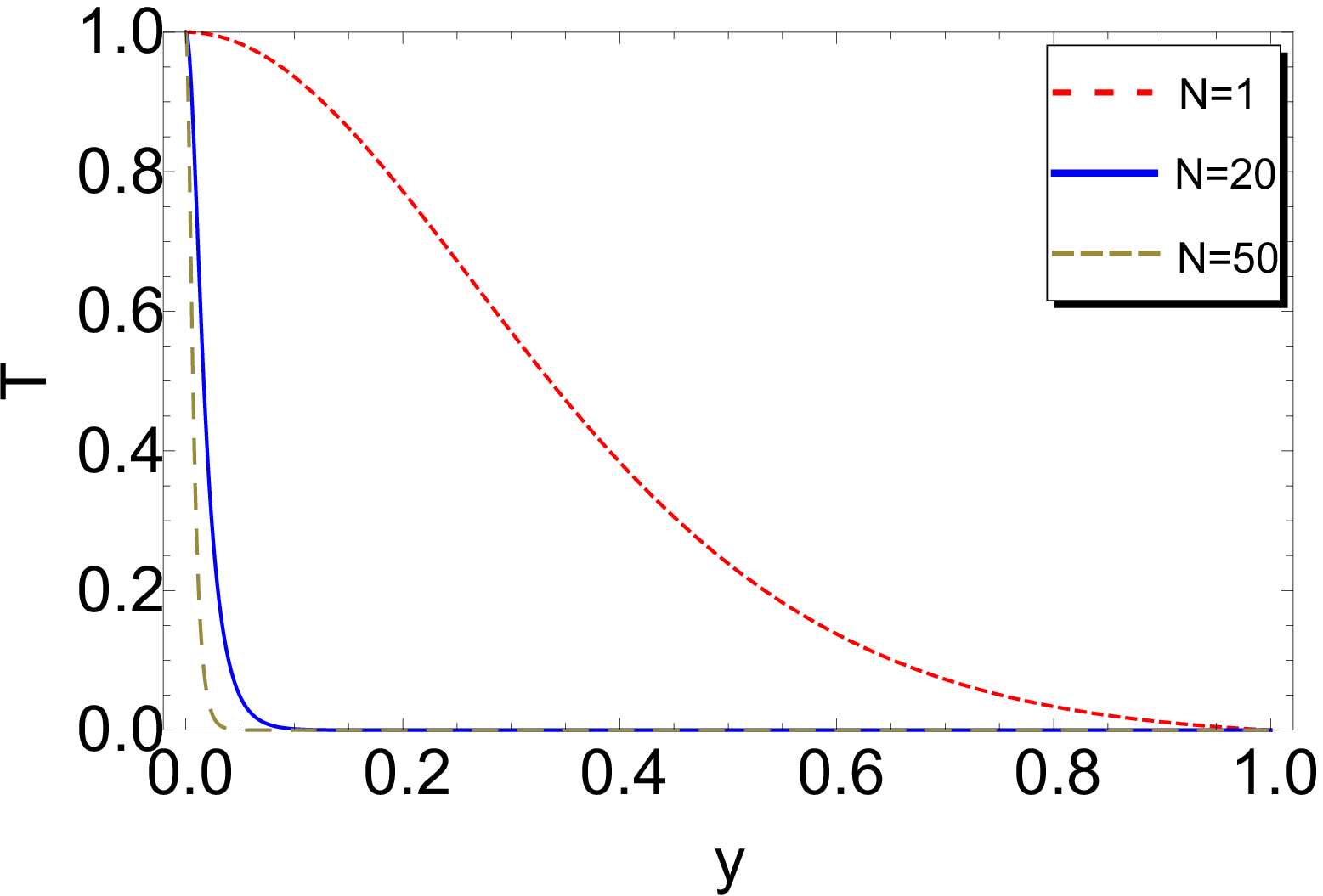}\begin{picture}(0,0)(0,0)
\put(-65,130){\makebox(0,0)[b]{$(b)$}}
\end{picture}\\
\caption{(Color online) The transmission probability $T_{N}$
as a function of the incident direction of the electron flow
$y=k_{y}/k_{F}$:  (a)$k_{F}L=0.1$; (b) $k_{F}L=1$.} \label{diod}
\end{figure}
If any of the parameters $L_{1}$, $L_{2}$, or $N$ are zero,
or the condition $kL_1=\pi n, \quad n=0,\pm 1, \dots$ is fulfilled,
Eq.(\ref{trans}) determines the transmission probability through the
sharp step: $T(k_{y})=\cos^{2}\varphi$.
Note, that in the n-p junction creating by the ripple-flat
graphene system, the Fermi momentum $k_{F}$ depends on the ripple radius
[see also Eq.(\ref{ef})]
\begin{equation}
E_F=\gamma k_F\Rightarrow k_{F}=\frac{|\alpha|}{2\gamma}\left(\frac{a}{R}\right)^{2}=0.46\left(\frac{a}{R}\right)^{2}nm^{-1}\,.
\label{kf}
\end{equation}
Evidently, one is able to control the degree of focusing of the electron beam by
fine turning of the angle $\phi$ with the aid of the discussed parameters and,
additionally, by means of the ripple radius as well.

 In contrast, for nonzero
values of the above parameters, we expect a smooth n-p junction.
 In order to trace  the dependence of the transmission probability
on the incident angle of electrons, we calculate numerically Eq.(\ref{trans})
at $L_{1}=L_{2}=L$ (see Fig.\ref{diod}).
It is noteworthy that the superlattice leads to the selective transmission of electrons.
For a small number of N elements in the ${\cal S}$ subsystem the transmission probability
is nonzero for a wide range of values of $k_y$ (see results for $N=1,20$).
However, the larger the number of N elements in the superlattice, the stronger
the selectivity effect for ballistic electrons. Our system focuses the electronic flow,
selecting the transmission of those trajectories that are close to the normal incidence.
In fact, for a large enough number N elements of the superlattice
the selection does not depend on the incident direction of an electron flow at all !
Indeed, at $N\gg 1$, only for the direction perpendicular to the surface of the ${\cal S}$ subsystem
there is almost the ideal transmission,
while for the other angles ($k_y\neq 0$) there is the strong backscattering.

\begin{figure}{}
\centerline{\includegraphics{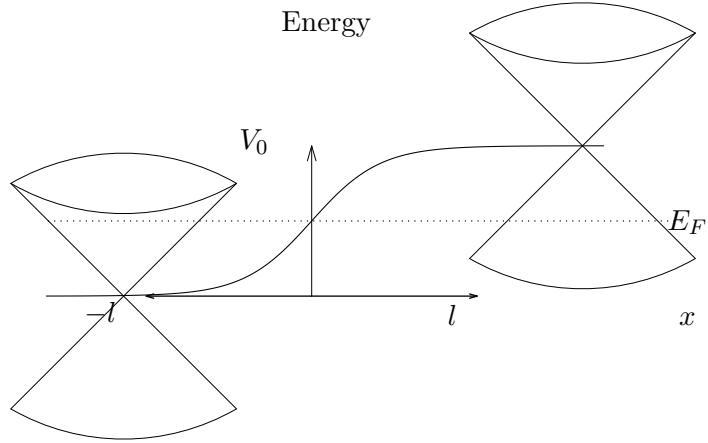}\unitlength=1.2mm\begin{picture}(0,0)(0,0)
\put(-49,32){\makebox(0,0)[b]{$V_{0}$}}
\put(-41,45){\makebox(0,0)[b]{Energy}}
\put(-1,23){\makebox(0,0)[b]{$E_{F}$}}
\put(-66,13){\makebox(0,0)[b]{$-l$}}
\put(-27,13){\makebox(0,0)[b]{$l$}}
\put(-1,13){\makebox(0,0)[b]{$x$}}
\end{picture}}
\caption{The sketch of the smooth interface potential.}
\label{exp}
\end{figure}

To  elucidate the advantage/disadvantage of the superlattice effect
for ballistic transport we compare the obtained results with those
obtained with the aid of the smooth step potential in the region around
the n-p junction
We model this by the Hamiltonian in the form
\begin{equation}
\hat{H}=v(x)\sigma_{0}+\gamma
(\hat{k}_{x}\sigma_{x}+\hat{k}_{y}\sigma_{y})\,.
\label{hr}
\end{equation}
Here, the smooth step potential
\begin{equation}
\label{st}
 v(x)=\frac{V_{0}}{2}\left( 1+\tanh(x/\ell))\right)
\end{equation}
is defined in the region $-\ell\leq x\leq \ell$ (see  Fig.\ref{exp}).
The transmission probability for the potential (\ref{st})  is
determined by the expression
\begin{equation}
\label{36}
T_{sm}(k_{y})=\left[\sinh\left(\pi
k_{F}\ell\sqrt{1-\left(\frac{k_{y}}{k_{F}} \right)^{2}} \right)/\sinh
\pi k_{F}\ell\right]^{2}\,.
\end{equation}
Details of calculations could be traced with the aid of Ref.\cite{Flu}.
In our case $E_F=\gamma k_F\equiv V_0/2$, and, correspondingly,  $k_{F}=V_{0}/2\gamma$.
In the limit $k_{F}\ell \ll 1$, we obtain the transmission
probability through the sharp step, $T(k_{y})=\cos^{2}\varphi$.
In contrast, at the condition $k_{F}\ell \gg 1$ we obtain (see Appendix \ref{ap1})
\begin{equation}
\label{36z} T_{sm}(k_{y}) \approx \exp(-\pi k_{F}\ell \sin^{2}\varphi)\,,
\end{equation}
which coincides with the result \cite{Falko}.
\begin{figure}[thb]
\includegraphics[height=6cm,clip=]{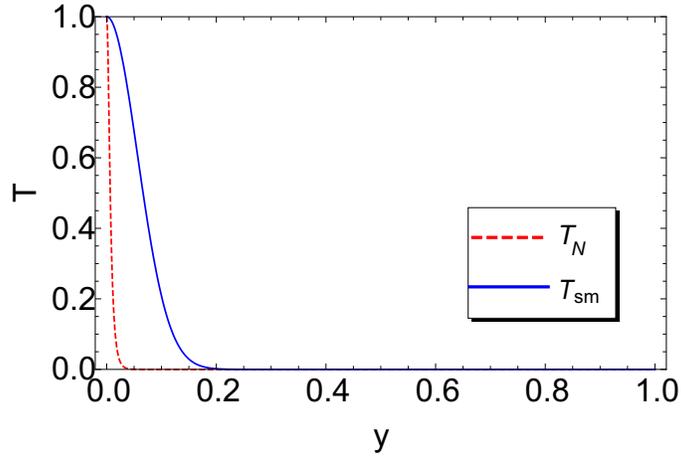}
\begin{picture}(0,0)(0,0)
\end{picture}\\
\caption{(Color online) The transmission probabilities $T$ as a
function $y=k_{y}/k_{F}$: dashed line connects the results for
$N=50$ elements ($T_{N}$) at $k_{F}L=1$; solid line connects the
results for the smooth potential ($T_{sm}$) at $k_{F}\ell=50$.}
\label{flugge}
\end{figure}
Thus, for the potential (\ref{st}) the range of transmitted angles is controlled
by the ratio $k_F\ell\sim\ell/\lambda_F$, where for our choise of parameters [see also Eq.(\ref{kf})]
\begin{equation}
\lambda_F\approx 13.65 \Bigg{(}\frac{R}{a}\Bigg{)}^2 nm\,.
\end{equation}
 The superlattice, that consists of
50 units, produces the selectivity that is much stronger than the one of
 the smooth potential step (see Fig.\ref{flugge}). In the both cases the
length of the interface is $100/k_{F}$.

Before to finalize this section there are a few comments in order.
Note that our results are valid for both $K$ and $K^{'}$ valleys. The basic difference between the
corresponding Hamiltonians consists in the sign of the momentum  $k_{y}$
(see for details \cite{Ando}).
While in our model the results depend on $k_{y}^{2}$ form, and, correspondingly, our conclusions
are valid for $K$ and $K^{'}$ valleys. We recall that  the translational invariance along the y axis
is kept in our model, while the edge effects are neglected.
Therefore, the average current is defined as
${\bf j}=s{\bf k}/k$ (see also Sec.2.4 in Ref.\cite{Allain}),
where $s=\sign(E_{\sl kin})$[$E_{\sl kin}\equiv s\gamma\sqrt{k_x^2+k_y^2})]$.

This is true only if one neglects the dependence  of the hopping integrals between
$\pi$ orbitals in
zig-zag or armchair graphene surface curvatures.
In order to illuminate the effect of this dependence on two interfaces, we have to calculate
the shift in the origin of $k_{x,y}$ by $\Delta k_{x,y}$,  produced by terms
of the order of $(a/R)^2$ neglected in our consideration.
In our analysis
we follow the arguments discussed by Ando (see $\S 5$ in \cite{Ando}).
In the effective mass approximation
for the zig-zag interface we obtain
\begin{equation}
\Delta k_{x}=\mp \frac{a}{4\sqrt{3}R^{2}}\left(1-\frac{3}{8}\frac{\gamma'}{\gamma}\right)\,,
\end{equation}
\begin{equation}
\Delta k_{y}=0\,,
\end{equation}
where the upper sign corresponds to the $K$ point, while the lower sign to the $K^{'}$ point. The parameter $\gamma=\sqrt{3}\gamma_{0}a/2
=-\sqrt{3}V_{pp}^{\pi}a/2$, $\gamma^{'}=\sqrt{3}(V_{pp}^{\sigma}-V_{pp}^{\pi})a/2$, where
$V_{pp}^{\pi}$ and $V_{pp}^{\sigma}$ are the hopping integrals for
$\pi$ and $\sigma$ orbitals, $a$ is the length of the primitive
translation vector.
We recall that in our model it is assumed that
that $V_{pp}^{\pi}\approx -3$ eV and $V_{pp}^{\sigma}\approx 5 $ eV.
 Therefore, we have $\gamma^{'}/\gamma \approx 8/3$, i.e., $\Delta k_{x}\approx 0$.
Thus, in the case of zig-zag interface the shifts are negligibly small, i.e.,
$\Delta k_{x}\approx 0, \quad \Delta k_{y}=0$. We conclude that the symmetry
between $K$ and $K^{'}$ valleys is conserved, Q.E.D.

In the case of the armchair interface we obtain that
\begin{equation}
\Delta k_{x}=0\,,
\end{equation}
\begin{equation}
\Delta k_{y}=\mp \frac{a}{4\sqrt{3}R^{2}}\left(\frac{5}{8}\frac{\gamma'}{\gamma}-1\right)\,.
\end{equation}
In other words, $\Delta k_{x}=0, \quad \Delta k_{y}\neq 0$.
In this case this curvature dependence  breaks the symmetry between $K$ and $K^{'}$ valleys.
The analysis of the effects related to this symmetry breaking requires the separate studies and
is beyond the scope of the present paper.

\subsection{Conductance}
From the transmission probability, the conductance is given by the Landauer formula
\begin{equation}
\label{72}
G_{N}=4\frac{e^{2}}{h}\int_{-k_{F}}^{k_{F}}T_{N}(k_{y})\frac{dk_{y}}{2\pi/W}=4\frac{e^{2}}{h}\frac{k_{F}W}{\pi}I_{N}\,.
\end{equation}
Here, the integral $I_{N}$, defined by the expression
\begin{equation}
\label{73} I_{N}=\int_{0}^{1}T_{N}(u)du, \ \ u=\frac{k_{y}}{k_{F}}\,,
\end{equation}
characterizes the efficiency of the
selection. For example, at $L_{1}=L_{2}=L$ and
$k_{F}L\approx 1$, we obtain
for $N=1,2$:
\begin{equation}
\label{74}
I_{1}=\int_{0}^{1}\frac{1-u^{2}}{\left(1+4(k_{F}L)^{2}u^{2}\right)^{2}}du
=0.33
\end{equation}
and
\begin{equation}
\label{75}
I_{2}=\int_{0}^{1}\frac{1-u^{2}}{\left(1+\left[16(k_{F}L)^{4}+12(k_{F}L)^{2}\right]u^{2}\right)^{2}}du
=0.14
\end{equation}
For the perfect transmission, i.e., for
$T(k_{y})=1$ the conductance
\begin{equation}
\label{72y}
G_{o}=4\frac{e^{2}}{h}\int_{-k_{F}}^{k_{F}}\frac{dk_{y}}{2\pi/W}=4\frac{e^{2}}{\pi
h}k_{F}W
\end{equation}
is the natural unit, since  $G_{N}=G_{o}I_{N}$.

\begin{figure}
\includegraphics[height=6cm,clip=]{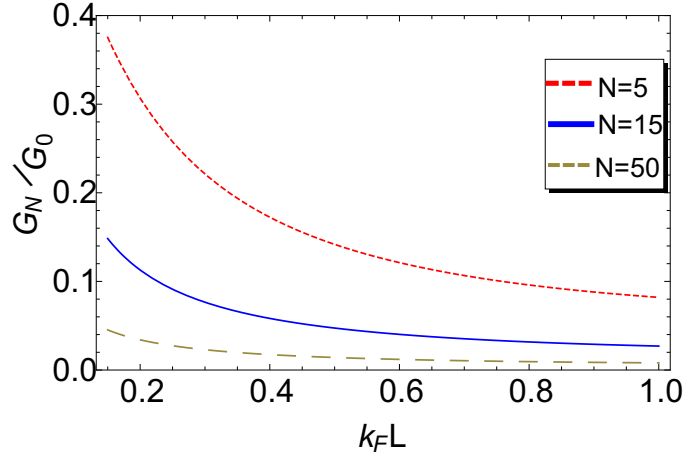}\begin{picture}(0,0)(0,0)
\end{picture}\\
\caption{(Color online) The relative conductance $G_{N}/G_{o}$ as a
function of the dimensionless parameter $k_{F}L$ for various values
of $N$ units in super-lattice section.} \label{int}
\end{figure}

The selective electrons transmission across the interface
created by $N$ units is demonstrated on Fig.\ref{int}, where the
dependance of $G_{N}/G_{o}$ on the dimensionless parameter $k_{F}L$
is depicted. The electron conductivity $G_{N}$ across the interface
with $N$ units is much smaller in comparison to $G_{o}$ for enough
large $N$.

We recall
that the estimation for the smooth step yields the value
\cite{Allain,Falko}
\begin{equation}
G_{sm}= 2\frac{e^{2}}{\pi
h}W\sqrt{\frac{k_{F}}{l}}=\frac{G_{o}}{2\sqrt{k_{F}\ell}}\,,
\end{equation}
that describes the  selectivity effect at the condition $k_{F}\ell\gg1$.

In order to achieve
the smooth step effect, the corrugations with gradually increasing
curvature can be used in our case. This conditions leads to
the inequality
\begin{equation}
G_{sm} > G_N\Rightarrow \sqrt{2\pi\ell/\lambda_F}\times I_N<1/2\,.
\end{equation}
If we hold fixed the condition $\ell=NL$, this inequality
determines the number of elements  $N$ and their length $L$
at the same length $\ell$ for the
smooth potential and the superlattice. Thus, by appropriate choice
of the product $NL$ one can always use the
advantage  of electron flow focusing through  the superlattice, which number of elements can be
controlled externally. Moreover, one can use additionally the fine turning of the ripple radius
and change carrier charge densities on different sides of our hybrid system.

\section{Summary}

Based on the fact of the different type of hybridization of carbon atom orbitals in
the flat and the corrugated graphene pieces, we developed the  model
of n-p junction. The $\pi$ orbital dependence on the surface
curvature means that the local chemical potential varies with the
curvature. In the approximation of the effective mass Hamiltonian, this fact
corresponds to the
effective electric field that depends on the electron position.
This effect becomes important once it would be possible to create a
graphene system with controlled variation of the surface curvature.
In fact, there are a few experimental techniques that demonstrate
evidently a spatial variation in graphene sheets nowadays.
For example, the electrostatic manipulation
allows to form ripples without any change of doping \cite{car19}.
 Another approach is based on the chemical vapor deposition that
provides quite promising way to create periodic nanoripples \cite{ni}.
It is found that ripples or wrinkles act as potential barriers for
charged carriers leading to their localization \cite{car16}.
This fact confirms our findings (see Sec.IIIA) and serves
as a solid argument in a favour of the vitality of our model.
Indeed, it is observed that the potential surface variations
are of around of 20-30 meV. In our model the ripple of the radius $R=12{\AA}$
yields the energy difference
between the flat and curved graphene pieces
$\Delta \varepsilon \approx 24 meV$ (see Eq.\ref{endif}).
The THz time-domain spectroscopy may be used to provide
contact-less, highly accurate information on conducting
properties of the curved graphene surface
(for a review see Ref.\cite{opt}), discussed
in our paper.

Our analysis of the hybrid system that consists of the
rippled +semiripple+flat pieces demonstrates the strong selectivity
effect of transmitted electron trajectories. The ballistic electron
transmission [see Eq.(\ref{trans})] depends on the radius of the
ripple, on the length of the arc of the ripple and on the width of
the flat region between ripples.   In fact, our system yields the
higher selectivity in contrast to the one produced by the smooth step
interface (see Fig.\ref{flugge}).
Most important, that the superlattice, described in the
paper, enables to one to control the conductance without
\textit{any additional electrical or magnetic sources}.
Namely, the selectivity is controlled by
the number of suitable N elements of the superlattice. The larger is the
number of elements N, the stronger is the selectivity.
At $N\gg 1$, only for the direction perpendicular to the surface of the ${\cal S}$ subsystem
there is almost the ideal transmission,
while for the other angles ($k_y\neq 0$) there is the strong reflection.
This phenomenon is due to the Klein tunneling that is grown in our system by virtue of
controlled graphene surface curvature.

\section*{Acknowledgments} The work was supported in part by Slovak Grant
Agency VEGA Grant 2/0009/19.

\appendix

\section{The smooth potential step}
\label{ap1}

The transmisson (\ref{36})
\begin{equation}
\label{36ap} T(k_{y})=\left[\sinh\left(\pi
k_{F}\ell\sqrt{1-\left(\frac{k_{y}}{k_{F}} \right)^{2}} \right)/\sinh
\pi k_{F}\ell\right]^{2}
\end{equation}
can be expressed in the limit $k_{F}\ell\gg 1$ as
\begin{equation}
\label{36x} T(k_{y})\approx \left[\exp\left(\pi
k_{F}l\sqrt{1-\left(\frac{k_{y}}{k_{F}} \right)^{2}} \right)/\exp
\pi k_{F}l\right]^{2}\,.
\end{equation}
Assuming that  nonzero transmission probabilities $T(k_{y})\neq 0$ exist only at
the condition $k_y/k_F \ll 1$, we have finally
\begin{equation}
\label{36y}
T(k_{y})\approx \left[\exp\pi
k_{F}l\left(1-\frac{1}{2}\left(\frac{k_{y}}{k_{F}}
\right)^{2}\right) /\exp \pi k_{F}l\right]^{2}\approx
\exp\left[-\pi k_{F}l
\left(\frac{k_{y}}{k_{F}} \right)^{2}\right]\,,
\end{equation}
which is the formula (\ref{36z}).

In the opposite limit $k_{F}\ell \ll 1$,  the transmission (\ref{36}) yields
the expression
\begin{equation}
\label{36xa} T(k_{y})\approx \left[\frac{\pi
k_{F}l\sqrt{1-\left(\frac{k_{y}}{k_{F}} \right)^{2}} }{\pi
k_{F}l}\right]^{2}\,,
\end{equation}
where it was used the approximation $\sinh\alpha \approx \alpha$ for $\alpha \ll 1$.
As a result we obtain
\begin{equation}
\label{36xb} T(k_{y})\approx 1-\left(\frac{k_{y}}{k_{F}}
\right)^{2}=\cos^{2}\varphi\,.
\end{equation}


\begin{thebibliography}{999}
\bibitem{Ando05}
             Ando T J
             2005 \textit{J. Phys. Soc. Jpn.} \textbf{74} 777.

\bibitem{13}
             Novoselov K S, Geim A K, Morozov S V, Jiang D, Katsnelson M I,
             Grigorieva I V, Dubonos S V and Firsov A A
             2005 \textit{Nature}   \textbf{438} 197.

\bibitem{14}
             Zhang Y, Tan Y-W, Stormer H and Kim P
             2005 \textit{Nature} \textbf{438} 201.

\bibitem{kat1}
             Katsnelson M I, Novoselov K S and Geim A K
             2006  \textit{Nat. Phys.} \textbf{2} 620.

\bibitem{Falko}
           Cheianov V V and Fal'ko V I
           2006 \textit{Phys. Rev. B} \textbf{74} 041403(R).


\bibitem{ch1}
            Cheianov V V, Fal'ko V I and Altshuler B L
            2007 \textit{Science} \textbf{315} 1252.

\bibitem{huard}
             Huard B, Sulpizio J A, Stander N, Todd K, Yang B and Goldhaber-Gordon D
             2007 \textit{Phys. Rev. Lett.} \textbf{98} 236803.

\bibitem{sutar}
             Sutar S, Comfort E S, Liu J, Taniguchi T, Watanabe K and Lee J U
             2012 \textit{Nano Letts.} \textbf{12} 4460.


\bibitem{mag}
             Chen S, Han Z, Elahi M M, Habib K M M, Wang L, Wen B, Gao Y,
             Taniguchi T, Watanabe K, Hone J, Ghosh A W and Dean C R
             2016 \textit{Science} \textbf{353} (6307) 1522.

\bibitem{klein}
             Klein O
             1929 \textit{Z. Phys. A: Hadr. Nucl.} \textbf{53} 157.

\bibitem{switch}
            Wang K, Elahi M M, Wang L, Habib K M M, Taniguchi T, Watanabe K,
            Hone J, Ghosh A W,  Lee G-H and Kim P
            2019 \textit{PNAS} \textbf{116} (14) 6575.

\bibitem{kumar}
           Kumar S and Parks D M
           2015 \textit{Nano Letts.} \textbf{15} 1503.

\bibitem{jul1}
          Baimova J A, Dmitriev S V, Zhou K and Savin A V
          2012 \textit{Phys. Rev. B} \textbf{86} 035427.

\bibitem{jul2}
          Liu B, Reddy C D, Jiang J, Baimova J A, Dmitriev S V,
          Nazarov A A and Zhou K
          2012 \textit{Appl. Phys. Lett.} \textbf{101} 211909.

\bibitem{Voz}
        de Juan F, Cortijo A and Vozmediano M A H
        2007 \textit{Phys. Rev. B} \textbf{76} 165409.

\bibitem{Kat}
        Katsnelson M and Geim A
        2008 \textit{Philos. Trans. R. Soc. A} \textbf{366} 195.

\bibitem{Guinea}
        Guinea F, Katsnelson M I and Vozmediano M A H
        2008 \textit{Phys. Rev. B} \textbf{77} 075422.

\bibitem{Allain}
        Allain P E and Fuchs J N
        2011 \textit{Eur. Phys. J. B} \textbf{83} 301.

\bibitem{PPCF}
        Pereira J M Jr, Peeters F M, Chaves A and Farias G A
        2010 \textit{Semicond. Sci. Technol.} \textbf{25} 033002.

\bibitem{PPN1}
        Pudlak M, Pichugin K N and Nazmitdinov R G
        2015 \textit{Phys. Rev. B} \textbf{92} 205432.

\bibitem{Ando}
         Ando T 2000
         \textit{J.Phys. Soc. Jpn.} \textbf{69} 1757.

\bibitem{PPN}
        Pichugin K N, Pudlak M and Nazmitdinov R G
        2014 \textit{Eur. Phys. J. B} \textbf{87} 124.

\bibitem{Wei}
        Wei Y, Wang B, Wu J, Yang R and Dunn M L
        2013 \textit{ Nano Lett.} \textbf{13} 26.

\bibitem{Bai}
         Bai Ke-Ke, Zhou Yu, Zheng H, Meng L, Peng H, Liu Z,  Nie J-C and He L
         2014  \textit{Phys. Rev. Lett.} \textbf{113} 086102.

\bibitem{low}
        Low T, Hong S, Appenzeller J, Datta S and Lundstrom M S
        2009 \textit{IEEE Trans. Elec. Dev.} \textbf{56} (6) 1292

\bibitem{Pudlak}
        Pudlak M and Pincak R
        2009  \textit{Eur. Phys. J. B} \textbf{67} 565.

\bibitem{Pudlak1}
        Pudlak M and Pincak R
        2009 \textit{Phys. Rev. A} \textbf{79} 033202.

\bibitem{SDD}
        Saito R, Dresselhaus G and Dresselhaus M S
        2003 \textit{Physical Properties of Carbon Nanotubes} (London: Imperial College Press).

\bibitem{Lomer}
          Lomer W M
         1955 \textit{Proc. Roy. Soc. A} \textbf{227} 330.

\bibitem{Yan}
         Yaniv A
        1978 \textit{Phys. Rev. B} \textbf{17} 3904.

\bibitem{Wallace}
        Wallace P R
        1947 \textit{Phys. Rev.} \textbf{71} 622.

\bibitem{Flu}
        Fl\H{u}gge S
        1994 \textit{Practical Quantum Mechanics} (Heidelberg: Springer-Verlag).

\bibitem{car19}
        Alyobi M M M, Barnett C J, Rees P and Cobley R J
        2019  \textit{Carbon} \textbf{143} 762.

\bibitem{ni}
        Ni G-X, Zheng Yi, Bae S, Kim H R, Pachoud A, Kim Y S, Tan C-L,
        Im D, Ahn J-H, Hong B H and \"{O}zyilmaz B
        2012 \textit{ACS Nano} \textbf{6} 1158.

\bibitem{car16}
        Vasi\'c B, Zurutuza A and Gaji\'c R
        2016  \textit{Carbon} \textbf{102} 304.

\bibitem{opt}
        B{\o}ggild P, Mackenzie D M A, Whelan P R, Petersen D H, Buron J D,
        Zurutuza A, Gallop J, Hao L and Jepsen P U
        2017 \textit{2D Mater.} \textbf{4} 042003.

\end{thebibliography}
\end{document}